# Percolative scale-free behavior in the boiling crisis

Limiao Zhang, Jee Hyun Seong, and Matteo Bucci*
*Department of Nuclear Science and Engineering*
*Massachusetts Institute of Technology, Cambridge, MA 02139, USA*

We present the first experimental observations of scale-free behavior in the bubble footprint distribution during the boiling crisis of water, in pool and flow boiling conditions. We formulate a continuum percolation model that elucidates how the scale-free behavior emerges from the near-wall stochastic interaction of bubbles and provides a criterion to predict the boiling crisis. It also offers useful insights on how to engineer surfaces that enhance the critical heat flux limit.

Boiling is a very efficient heat transfer process, widely applied for heat management, e.g., in electric power stations and high-power-density electronic devices. In such systems, the boiling process dynamics is driven by the heat flux transferred from the heated surface. An increase in the heat flux produces a rise of the surface temperature, which in turn increases the bubble nucleation site density and departure frequency. The chief vulnerability of boiling is an instability known as the boiling crisis, triggered when the heat flux reaches the critical heat flux (CHF) limit. This phenomenon coincides with a sudden transition from a nucleate boiling regime, with discrete bubbles on the surface, to a film boiling regime, where a stable vapor layer blankets the entire heating surface [1]. Such layer causes a drastic degradation of the heat removal process, resulting in a potentially catastrophic escalation of the heater temperature. Thus, understanding the boiling crisis, and predicting and possibly enhancing the CHF are desirable goals for the safety and economics of many thermal systems.

Experimental investigations have revealed that CHF depends on fluid properties and operating conditions, heater geometry, surface material, orientation, and properties (e.g., roughness, porosity, and intrinsic wettability). Many mechanisms and models have been proposed to capture these effects [2]. However, there is still no agreement within the thermal science community on the actual trigger mechanism for the boiling crisis, let alone a universal model to predict it. Historically, most models have been built assuming that the boiling crisis is triggered by a macroscale hydrodynamic instability in the countercurrent vapor/liquid flow far from the heating surface [3,4]. More recently, several authors have argued that the boiling crisis is instead a near-wall phenomenon, determined by micro-scale fluid-solid interactions on the heating surface [5-9]. However, while most models attempt to capture the CHF limits leveraging scale-based descriptions of presumed trigger mechanisms, recent findings suggest that the boiling crisis is a scale-free, critical phenomenon. This perception stems from observations of power-law spectra in the temperature fluctuations of wire heaters [10], energy distributions of acoustic emissions during surface quenching with liquid nitrogen [11], and bubble size distributions in slowed-down boiling of hydrogen at reduced gravity [12]. The predictions of a numerical lattice spin model presented in Ref. [11] suggest that the critical behavior may be the result of a bubble percolation process. The observations in Ref. [12] seem to corroborate this hypothesis. However, one cannot exclude that these findings are affected by the special dynamics of the boiling process in such operating conditions (e.g., the bubble departure diameter tends to infinity) [12]. To affirm the percolative scale-free nature of the boiling crisis, an experimental and theoretical demonstration in conditions of broad relevance, such as the pool and flow boiling of water, is still unavailable and clearly necessary.

In this letter, we present the first experimental study of bubble footprint distributions during the pool and flow boiling of water. At CHF, the experimental distributions follow a power-law $1/A^\gamma$ with a critical exponent $\gamma$ smaller than 3, which demonstrates the scale-free nature of the boiling crisis [13]. Our experiment also enables measurement of fundamental boiling parameters (i.e., nucleation site density, growth time, bubble departure frequency and radius), and is instrumental in revealing the dynamics of the bubble interaction process. Inspired by these observations, we develop a bubble percolation model based on the continuum percolation theory [14]. The model explains and captures how the scale-free behavior at the boiling crisis emerges from the near-wall stochastic interaction of individual bubbles. It also provides a criterion to predict the boiling crisis.

We run boiling experiments featuring specially-designed heaters (see schematic in Fig. 1) consisting of a 1 mm thick, infrared (IR) transparent sapphire substrate coated on one side with an electrically conductive, IR opaque, 0.7 µm thick layer of Indium-doped Tin Oxide (ITO). The thin ITO coating, in contact with water, is the Joule heating element. It has negligible thermal resistance and heat capacity, i.e., the ITO temperature coincides with the actual temperature on the boiling surface. The heater is installed in a pool boiling or a flow boiling apparatus, as detailed in the Supplemental Material, Sec. 1 [15]. During the experiments, we increase the heat flux released by the ITO in a sequence of steady steps, up to the critical heat flux that causes the boiling crisis. For each heat flux, the infrared radiation emitted by the ITO and transmitted through sapphire is recorded by a high-speed infrared camera (with a temporal resolution of 400 μs and a spatial resolution of 115 μm/pixel) and post-processed to obtain the time-dependent temperature and heat flux distributions on the boiling surface [16]. The high temporal and spatial resolution enabled by this technique is key to capturing the dynamics of the boiling process up to the boiling crisis. Bubbles nucleate, grow, perhaps coalesce, and depart from the surface. The life of discrete bubbles is described through finite, characteristic parameters, i.e., departure frequency, growth time, and departure diameter. Bubbles interaction is instead a stochastic process determined by these parameters, as well as the distance among the nucleation sites, i.e., the nucleation site density. A typical output is shown in Fig. 1, where we can clearly distinguish the patches of individual and coalesced bubbles attached to the surface.

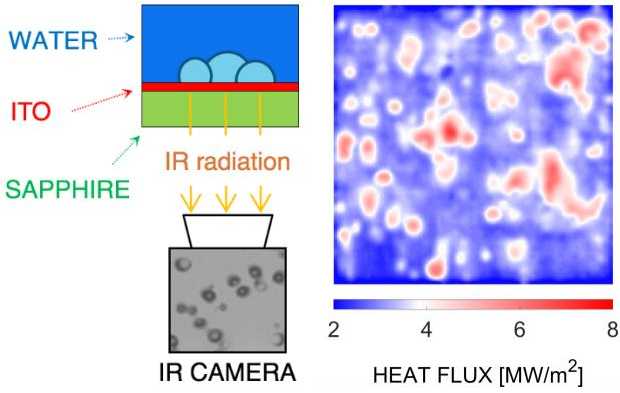

FIG. 1. The schematic on the left shows the ITO-sapphire heater configuration with the IR diagnostics (not to scale). The image on the right shows a typical instantaneous heat flux distribution on the boiling surface (1x1 cm$^2$, flow boiling test at atmospheric pressure, bulk temperature 95 °C, mass flux 2000 kg/m$^2$/s and heat flux 3460 kW/m$^2$).

These distributions are processed to measure nucleation site density, bubble wait time and growth time and, importantly, the footprint area of each bubble on the boiling surface, using the techniques described in Ref. [17]. The measured bubble footprint area probability density functions (PDFs) for saturated pool boiling and subcooled flow boiling tests run at atmospheric pressure are shown in Fig. 2 (a) and (c) (for the flow boiling tests the bulk temperature is 95 °C, i.e., ~5 °C below the saturation temperature at the system pressure, the mass flux is 2000 kg/m$^2$/s). For low heat fluxes (e.g., at 300 kW/m$^2$ and 1270 kW/m$^2$ for the pool boiling and flow boiling experiments, respectively), most of the bubbles are isolated, i.e., they seldom interact with each other. We observe that, under these conditions, the footprint area PDF is exponentially damped (as also observed in Ref. [12]), and can be fitted by

$$P(A) = ce^{-cA}. \qquad (1)$$

Assuming that the footprint of individual bubbles is circular, the bubble footprint radius PDF is

$$P(R) = 2\pi Rce^{-c\pi R^2}. \qquad (2)$$

Note that from Eq. (2), $c$ is related to the average bubble footprint radius $\langle R \rangle$ ($= 1/\sqrt{4c}$). As the heat flux increases (e.g., at 800 kW/m$^2$ and 2580 kW/m$^2$ for the pool boiling and flow boiling experiments, respectively), bubbles merge more and more frequently, and large bubble patches start to appear on the boiling surface. When the boiling crisis occurs (at 1170 kW/m$^2$ and 4320 kW/m$^2$ for the pool boiling and flow boiling experiments, respectively), the bubble footprint area PDF is a power law function $1/A^\gamma$. In pool boiling (Fig. 2 (a)), the critical exponent $\gamma$ evaluated by the maximum likelihood method and tested by the Kolmogorov-Smirnoff method [18,19] is $1.52 \pm 0.01$. The critical exponent for the flow boiling test (Fig. 2 (c)) is instead $1.85 \pm 0.01$.

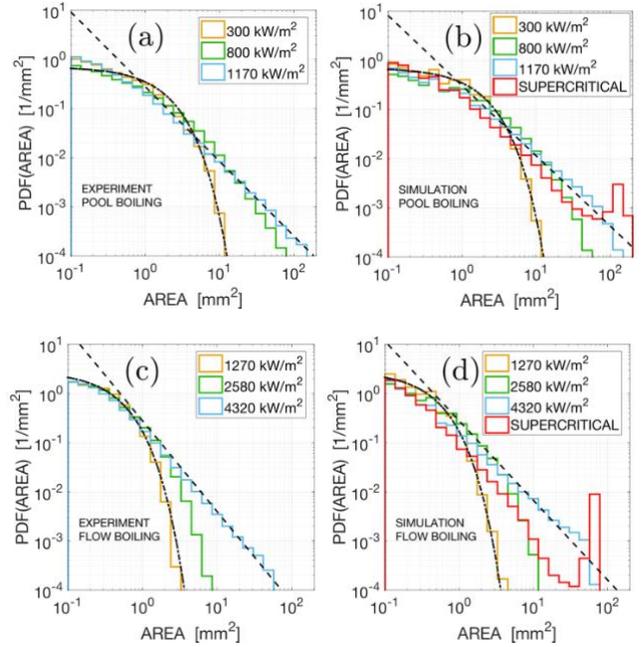

FIG. 2 (a) Experimental and (b) simulated bubble footprint area PDFs at different heat fluxes, from nucleate boiling (orange, 300 kW/m$^2$) till the boiling crisis (sky-blue, 1170 kW/m$^2$) for the saturated pool boiling tests at atmospheric pressure. The dashed-dotted line represents Eq. (1) obtained with an average bubble radius $\langle R \rangle$ = 0.6 mm. The dashed line represents a power law distribution $1/A^\gamma$. The power law exponents are $\gamma = 1.52 \pm 0.01$ and $\gamma = 1.43 \pm 0.01$ for the experimental and the simulated PDFs, respectively. The red curve in (b) corresponds to a supercritical case which represents the film boiling regime. (c) and (d) are analogous experimental and simulated results for the subcooled flow boiling test. Here, the average bubble radius for the exponential fitting, i.e., Eq.(1), is $\langle R \rangle$ = 0.3 mm. The power law critical exponents are $\gamma = 1.85 \pm 0.01$ and $\gamma = 1.62 \pm 0.01$ for the experimental and the simulated PDFs, respectively.

The emergence of scale-free behavior at CHF indicates the absence of a characteristic scale in the boiling crisis, also in the pool and flow boiling of water. The different values of $\gamma$ suggest that the critical exponent depends on the operating conditions, which seems reasonable given the different boiling phenomenology (e.g., different forces affecting bubble departure dynamics) in pool boiling and flow boiling.

Inspired by our experimental observations, we introduce a Monte Carlo (MC) model based on the continuum percolation theory [14], which captures the footprint area PDFs. From our experiments, we know the nucleation site density $N''$, the bubble growth time $t_g$, and the bubble departure frequency $f$ for each heat flux. Given a surface of area $A$ equal to our heater active area (i.e., 1x1 cm$^2$), we randomly generate $N''A$ nucleation sites. The probability to have a bubble growing out of a certain nucleation site is equal to $t_g f$. Thus, given a random number $b \in [0,1]$, if $b \leq t_g f$, we generate a bubble, otherwise, we move to the next nucleation site. Similarly, if a nucleation site is already covered by a bubble growing out of another nucleation site, we move to the next one. When generating a bubble, its radius is determined according to the distribution of individual bubbles, Eq. (2), measured experimentally (for the derivation details see the Supplemental Material, Sec. 2 [15]):

$$R = \sqrt{-\frac{4\langle R \rangle^2}{\pi}\ln(1-q)}, \qquad (3)$$

where $q \in [0,1]$ is a random number. The average bubble radius $\langle R \rangle$ is the last characteristic quantity obtained from the experiments. However, the average size of individual bubbles can only be accurately measured at low heat fluxes, i.e., when bubbles do not interact with each other. At high heat fluxes, the individual bubble footprint area PDF shrinks compared to the distribution that bubbles would have if they were growing without ever interacting. To overcome this limitation, the value of $\langle R \rangle$ at high heat flux is estimated scaling the value of $\langle R \rangle$ at low heat fluxes by a bubble departure, force balance model [20] (for the derivation details and a complete description of the experimental data used as input in the MC model, see the Supplemental Material, Sec. 2 [15]). We repeat this procedure throughout the nucleation sites. A typical output image from one cycle is shown in Fig. 3, where we sample the areas of the bubble patches, for both individual and coalesced bubbles. Then, the process is repeated many times, until we obtain a converged bubble footprint area PDF.

The predicted PDFs are shown in Fig. 2 (b) and (d). The MC model captures the trends of the experimental PDFs correctly, at any heat flux level. The PDF at CHF follows a power law, which corroborates the thesis of the boiling crisis as a percolative near-wall phenomenon. Note that, despite the simplicity of the model, the simulated critical exponents are similar to the ones measured experimentally.

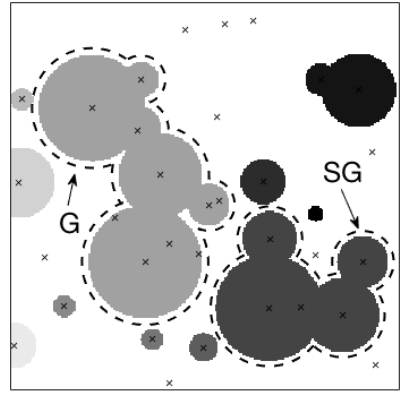

FIG. 3 Typical output of a MC iteration, with the identification of the giant $G$ and the second giant $SG$ clusters. Bubble footprints of the same color belong to the same cluster. Small crosses indicate nucleation sites.

The scale-free behavior is typical in critical phenomena [21], including percolative processes, such as forest fires [22] or traffic jams [23]. Percolation is a powerful tool to analyze phase transition in stochastic processes, with multiple applications in natural and engineering sciences [24]. Here, we show how continuum percolation can be used to predict the boiling crisis, assuming that we know how nucleation site density, bubble growth time and departure frequency, and bubble radius change with the heat flux, i.e., the boiling driving force. In classic site percolation models, a site on a lattice is occupied with probability $p$. Two neighbor nodes are connected if they are both occupied. A cluster is defined as a group of sites connected by near-neighbor distances [25]. There is a percolation threshold $p_c$ below which only a few isolated clusters exist, and the size of these clusters increases with $p$. Conversely, for $p$ above $p_c$, a single, large cluster percolates through the lattice. The emergence of a spanning cluster indicates the occurrence of percolation transition. The percolation threshold $p_c$ coincides with a maximum of the second largest cluster size, which provides the system failure criterion, and a scale-free cluster size distribution [26]. In continuum percolation, instead, the discrete lattice sites are replaced by penetrable objects (e.g., bubble footprints) in a continuous space (e.g., the boiling surface). The percolation threshold is given by a critical filling factor $\eta_c$, which plays a role similar to $p_c$ in site percolation [27].

In our work, we sample the area of the giant $G$ and the second giant $SG$ cluster (see Fig. 3), i.e., the bubble patches with the largest and second largest footprint area, respectively. The trends of the $G$ and $SG$ area, both measured (solid line) and simulated (filled dots), are plotted in Fig. 4 as a function of the nucleation site density, which monotonically increases with the heat flux. When the nucleation site density is small, for low heat fluxes, bubbles do not coalesce and the sizes of the $G$ and $SG$ clusters are very close, and are consistent with the size of individual bubbles.

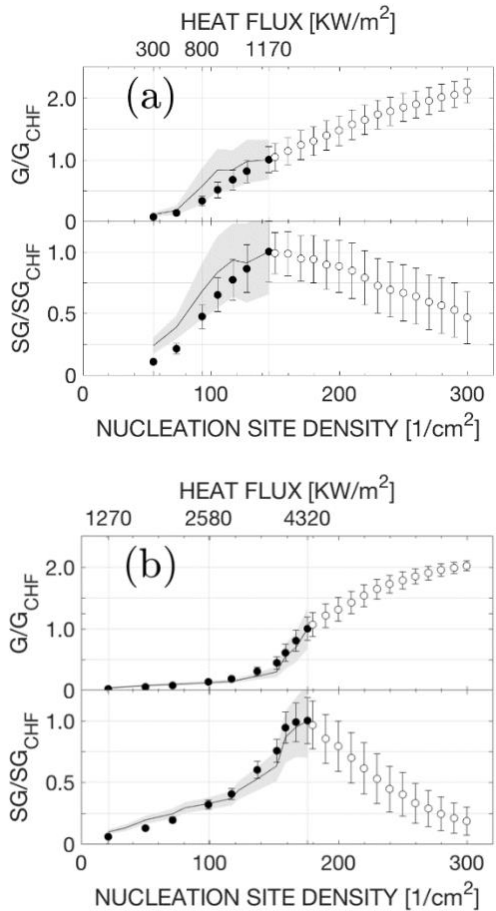

FIG. 4 Measured (solid line) and simulated (filled or empty dots) trends of the $G$ and $SG$ area as a function of the nucleation site density in pool boiling (a) and flow boiling conditions (b).

Then, as the heat flux increases, the growth of the clusters is first determined by the attachment of discrete bubbles. However, as the clusters grow bigger (i.e., for higher and higher heat fluxes), they start to coalesce with each other, and the larger is a cluster, the higher is the probability to absorb smaller ones. Thus, the $G$ and $SG$ areas increase with the nucleation site density up to the $N''$ corresponding to the boiling crisis (i.e., 145 sites/cm$^2$ for saturated pool boiling test and 176 sites/cm$^2$ for the flow boiling test). Right before the CHF, the $G$ area is still increasing, while the $SG$ area seems to reach a maximum. Note that, as the boiling crisis is detected, experiments are promptly interrupted to avoid the burnout of our heating element. Thus, the "experiment" is continued numerically by adding extra nucleation sites to the simulations. The results (hollow dots in Fig. 4) confirm that the boiling crisis coincides with a maximum of the $SG$ cluster area. It indicates a critical point, beyond which the process become unstable, as the largest cluster $G$ rapidly absorbs all the other clusters, i.e., there is a rapidly growing vapor layer, covering the entire boiling surface. The red curves in Fig. 2 (b) and (d) correspond to the simulated bubble footprint area PDFs in supercritical conditions ($N''$ = 300 sites/cm$^2$ for both pool boiling and flow boiling conditions), showing the emergence of a large spanning vapor patch. Note that, the maximum in the $SG$ size coincides with the occurrence of a power-law distribution, as expected in percolation criticality [26]. It provides a criterion to predict the boiling crisis and, considering the analogy with the dynamics of other percolative processes [26], further demonstrates the percolative, critical nature of the boiling crisis.

In conclusion, our investigation of pool and flow boiling of water (i.e., conditions of broad interest), reveals that the boiling crisis is a percolative, scale-free phenomenon. We propose a continuum percolation model and a criterion to predict the boiling crisis based on three parameters: $N''$, $t_g f$, and $\langle R \rangle$. The model decrypts the relationship between the driving force (i.e., the heat flux), the response (i.e., nucleation site density, growth time, bubble departure frequency and radius), and the stability of the process (i.e., the boiling crisis). The percolation threshold is defined by the "triplet" $N''$, $t_g f$, $\langle R \rangle$ that maximizes the $SG$ area, i.e., the boiling crisis is triggered by a "critical triplet". This finding has potentially profound practical implications, as it suggests that the boiling crisis can be delayed (i.e., CHF can be enhanced) by modifying the system response to the driving force. Assuming that we can tune these three parameters, e.g., by engineering the size and the number of the nucleation cavities, this criterion can help identifying optimal strategies to maximize the CHF limit. While our results are obtained on randomly positioned nucleation sites, our analysis suggests that the boiling crisis could be delayed by constraining the position of the bubbles according to an optimized lattice and nucleation cavity size that maximize the number of nucleation sites, while limiting the bubble interaction probability.

We thank J. Buongiorno, E. Baglietto, S. Yip and A. Kossolapov for many insightful comments. L. Z. acknowledges the Chinese Scholarship Council (CSC, 201706020179). J. H. S. acknowledges the Consortium for Advanced Simulations of Light Water Reactors (CASL).

*mbucci@mit.edu

# Percolative scale-free behavior in the boiling crisis: Supplemental Material

Limiao Zhang, Jee Hyun Seong, and Matteo Bucci

*Department of Nuclear Science and Engineering*

*Massachusetts Institute of Technology, Cambridge, MA 02139, USA*

## 1. Experimental setups and diagnostics

The implementation of our experiment requires special heaters, high-speed infrared (IR) diagnostics, and fairly complex image processing tools. Hereafter we provide a short description of the infrared heater, and the pool and flow boiling devices. Details about the postprocessing algorithms can be found in Ref. [2]. A summary of the measured boiling quantities used as inputs in the stochastic model is reported in Sec. 2.

*Infrared heater*

Our heaters consist of a 1-mm thick, IR transparent sapphire substrates (20 x 20 mm$^2$), coated with a 0.7 µm thick, IR opaque, electrically-conductive Indium Tin Oxide (ITO) coating (see Figure S1.1). The thermal response of the heater is determined by the sapphire substrate, which has thermal properties (e.g., thermal diffusivity and effusivity) very similar to Stainless Steel, Inconel, or Zircaloy [2]. The ITO coating, in contact with water, is the joule heating element. It has an electrical sheet resistance of 2.5 ohms per square. It has negligible thermal capacity or thermal resistance, i.e., the temperature gradient through the ITO thickness is negligible. Silver pads for electrical connections are deposited on top of the ITO and wrap around the filleted edges of the sapphire substrate. They limit the active ITO area to a 10×10 mm$^2$ square and allow a uniform release of electric power by the active ITO surface.

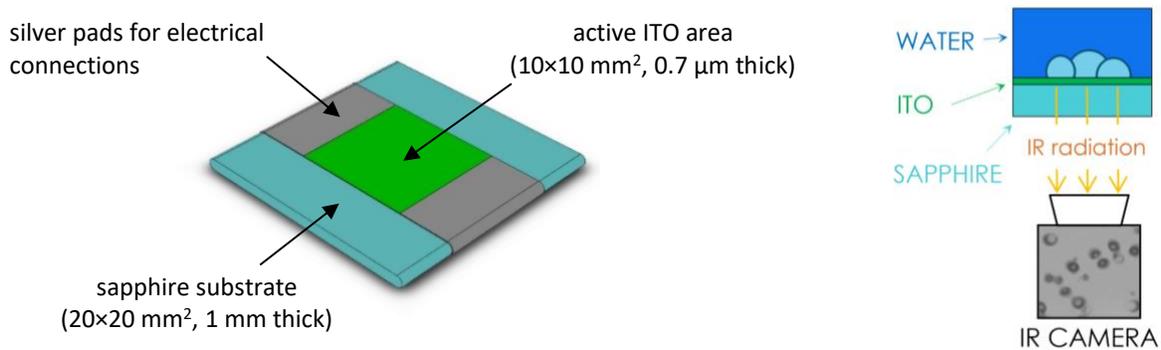

Figure S1.1. Left: sketch of IR heater (20×20 mm$^2$, 1 mm thick) consisting of a sapphire substrate, a 700 nm thick, electrically-conductive, IR-opaque ITO coating acting as Joule heater, and silver pads for electrical connections. Right: Schematic of the ITO-sapphire heater configuration with the IR diagnostics (not to scale).

The ITO coating is nano-smooth. Sapphire is also nano-smooth but has localized micro-scale imperfections (see Figure S1.2). These imperfections are conformally coated by the ITO (0.7 µm thick) and serve as bubble nucleation sites.

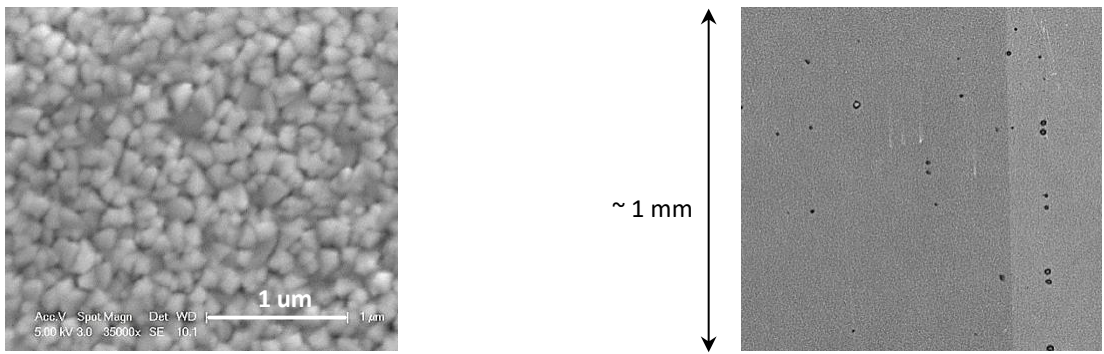

Figure S1.2. Pictures of the ITO surface with different magnification. The left figure is taken with an SEM and reveals a granular, nano-smooth surface. The right picture has been taken with an optical microscope: cavities that might act as nucleation sites appear as black dots or rings.

The ITO coating is opaque in the 3-5 µm wavelength range. Sapphire is instead quasi-transparent, and transmit almost entirely the IR radiation emitted by the ITO. This allows an IR camera to record the IR radiation emitted by ITO coating (IRC806HS, operated with a spatial and temporal resolution of 115 µm/pixel and 400 µs, respectively). However, due to the slightly absorbing nature of sapphire, the post-processing of the IR images requires the solution of a coupled conduction-radiation inverse problem, as discussed in Ref. [1]. Examples of temperature and heat flux distributions obtained by this technique can be found in the videos of Ref. [2].

*Pool boiling and flow boiling setups*

The test section used for pool boiling experiments is sketched in Figure S1.3. The pool boiling cell features a concentric-double-cylinder structure made of 316L stainless steel. Boiling of DI water at atmospheric pressure (e.g., ~1.01 bar) takes place in the inner cell, while the outer enclosure functions as an isothermal bath. The whole facility is surrounded by thermal insulating foam. The DI water in the inner cell is maintained in saturation conditions (i.e., at ~100 °C) by circulating a temperature-controlled fluid through the isothermal bath. The infrared heater is installed in a ceramic, Shapal$^{TM}$ cartridge, and sits at the bottom of the cell. The high-speed infrared camera IRC806HS visualizes the boiling surface from the bottom, using a hot mirror that reflects the infrared radiation emitted from the heater.

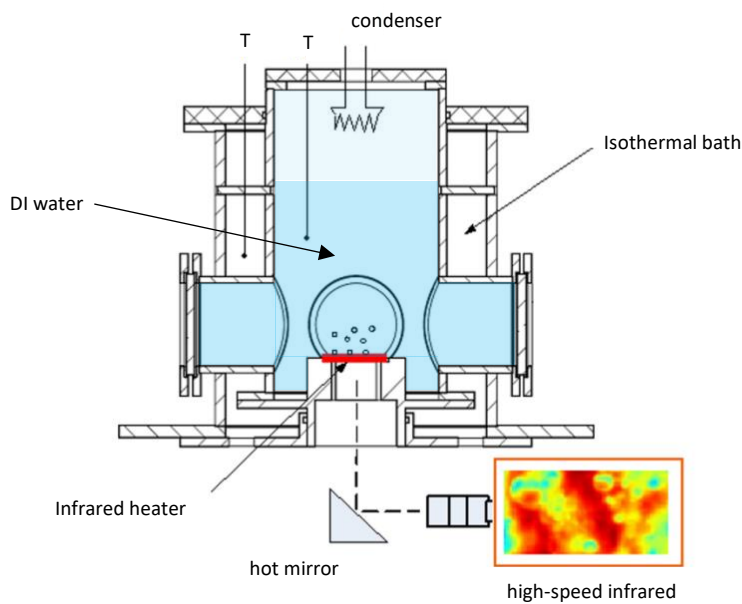

Figure S1.3. Schematic of the pool boiling with optical arrangement.

The flow boiling loop is built with stainless steel 316L and is designed for operation with deionized (DI) water (see Figure S1.4, Left). It is equipped with variable frequency pump, flow meter, temperature and pressure instrumentation, preheater, flow channel with the test section, chiller, accumulator and a fill and drain tank. Filtering and dissolved oxygen monitoring is accomplished via a secondary loop used during the initial stages of testing. A pump provides the requisite head for mass fluxes up to 2000 kg/m$^2$s, which was used in this experiment. The bulk temperature of the fluid is controlled by adjusting the power of the preheater and the secondary flow in the chiller. The present experiment was run at atmospheric pressure with slightly subcooled water at 95 °C. At the heart of the facility is a test section (see Figure S1.4, Right) with a $3 \times 1$ cm$^2$ flow channel running the length of the structure. It connects to an entrance region providing more than 60 L/D to establish fully-developed turbulent flow at the position of heater. The main body of the test section consists of four sides. Three are used for quartz windows to provide optical access; the fourth wall contains a ceramic cartridge, made of Shapal, used to hold the IR heater perfectly flushed with the channel walls (see Figure S1.5). The infrared radiation emitted by the heater is reflected towards the infrared camera by an infrared, hot mirror.

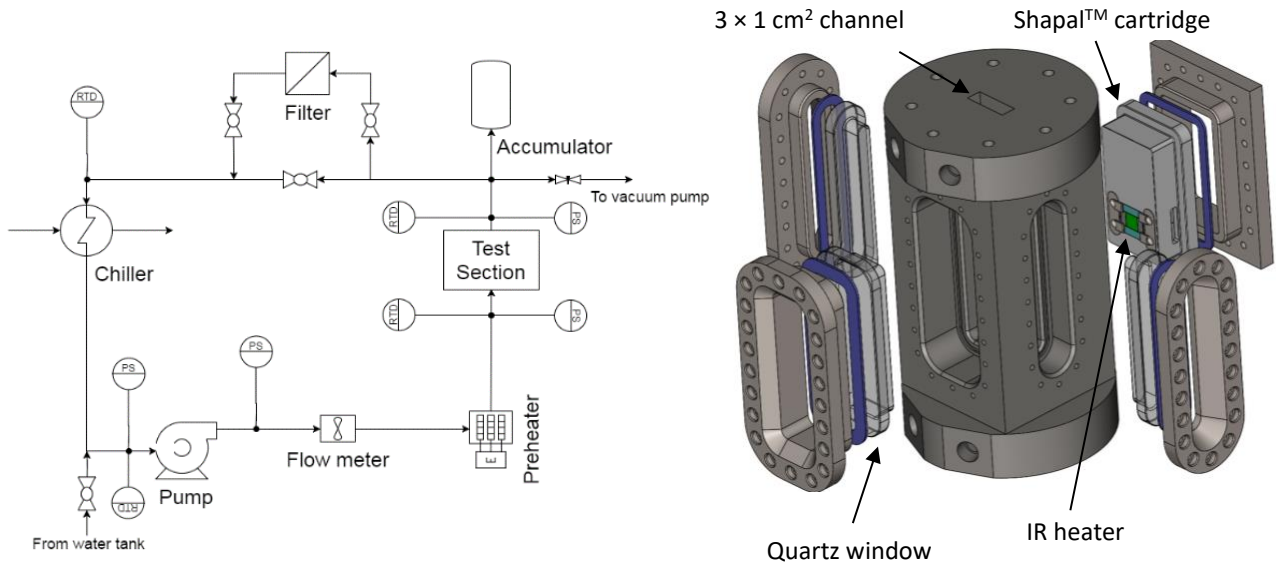

Figure S1.4. Schematic of the flow loop (left) and exploded view of the flow boiling test section (right) (adapted from [2]).

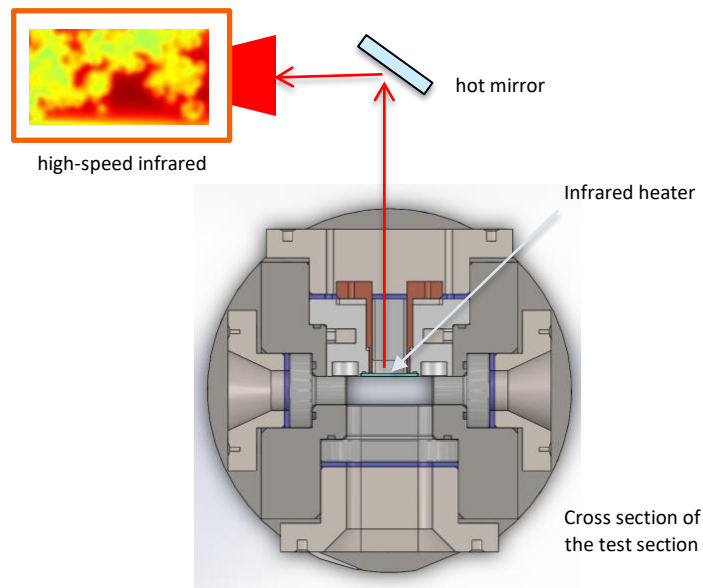

Figure S1.5. Cut-view of the test section with optical arrangement (adapted from [2]).

## 2. Experimental results

In this section, we report the measured boiling curves (Figure S2.1) and fundamental boiling quantities (Figure S2.2 through S2.4) used as inputs in the Monte Carlo (MC) model. Red dots and blue dots stand for pool boiling and flow boiling conditions, respectively.

The boiling curves (Figure S2.1) are obtained by time-averaging the heat flux and temperature distributions on the boiling surface. The boiling crisis occurs at the highest heat flux value for each curve, i.e., the critical heat flux (CHF). The heat flux is our experimental control parameter, and it is typically very stable. Its error bars do not extend beyond the markers (i.e., the dots) surface. The fluctuations of the average temperature are also very small, typically below 0.2 °C, and are also covered by the markers. These boiling curves also include experimental points recorded before the onset of nucleate boiling. They are shown as color-filled dots and correspond to conditions where the heat transfer regime is single-phase natural convection (for the pool boiling curve) or forced convection (for the flow boiling curve).

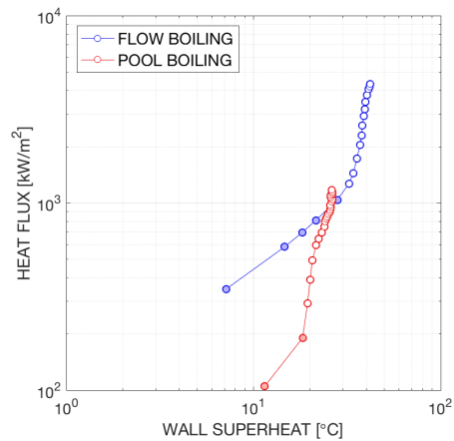

Figure S2.1. Boiling curves (time-averaged heat flux vs. time-averaged wall superheat) in flow boiling (blue) and pool boiling (red) experiments. Colored dots correspond to single phase heat transfer conditions, i.e., before the onset of boiling. Empty dots correspond to boiling heat transfer conditions.

The nucleation site density is plotted as a function of the average heat flux (Figure S2.2, Left) and the average wall superheat (Figure S2.2, Right). The detection of nucleation sites is quite accurate, and results in uncertainties which are typically within ± 3 nucleation sites per square centimeter even at high heat fluxes. Note that the wall superheat required to reach a certain nucleation site density is much higher in the flow boiling than in the pool boiling experiments. This difference can be attributed to the presence of the flow itself and the fact that the bulk water temperature is slightly colder (95 °C instead of 100°C).

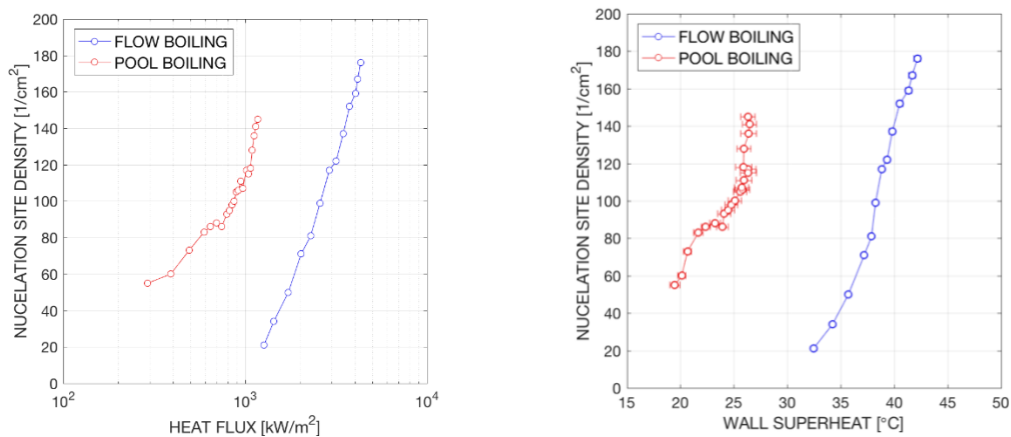

Figure S2.2 Nucleation site density as a function of average heat flux (Left) and average wall superheat (Right) in flow boiling (blue) and pool boiling (red) experiments.

Figure S2.3 shows the measured values of the product $t_g f$, i.e., the product of the bubble growth time, $t_g$, and the bubble departure frequency, $f$. Note that the bubble departure frequency is given by $f = 1/(t_w + t_g)$, where $t_w$ is the bubble wait time. De facto, $t_g f$ is the probability to find a bubble growing out of a certain nucleation site on the boiling surface. When heat flux and wall temperature increase, the bubble growth time increases, and the bubble wait time decreases. Accordingly, $t_g f$ increases with the heat flux and wall superheat. In this plot, the error bars are determined by the variance of the bubble probability, $t_g f$, for all the active nucleation sites.

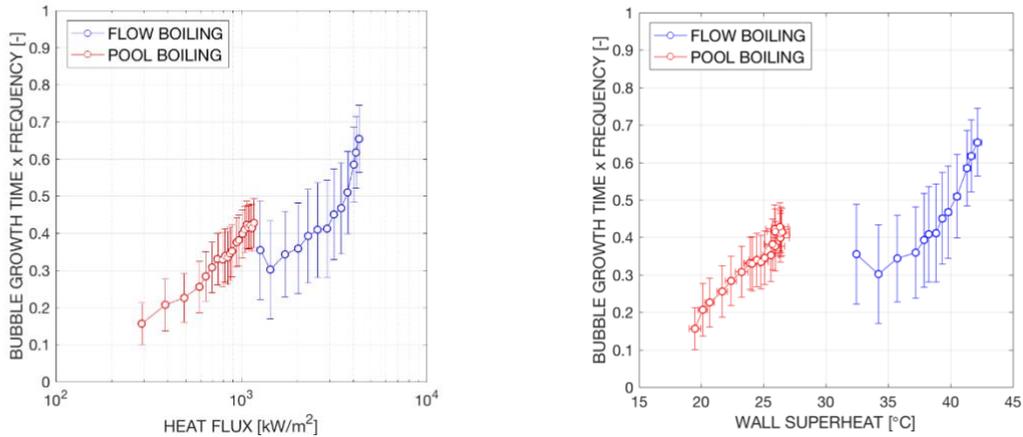

Figure S2.3. Product of bubble growth time and bubble departure frequency as a function of average heat flux (Left) and average wall superheat (Right) in flow boiling (blue) and pool boiling (red) experiments.

Figure S2.4 shows the bubble footprint radius of individual bubbles (empty dots), which can be accurately measured only at low heat fluxes. Under such conditions, most of the bubbles are isolated, i.e., they seldom interact with each other. Conversely, at high heat fluxes, the bubbles interact with each other, and the probability of interaction increases with the bubble size, i.e., with the heat flux. Therefore, data for these conditions are not shown; It is difficult to experimentally evaluate the radius distribution that bubbles would have if they were growing without ever interacting. To overcome this problem, the size of the bubbles at high heat fluxes is estimated by a mechanistic scaling model, discussed hereafter.

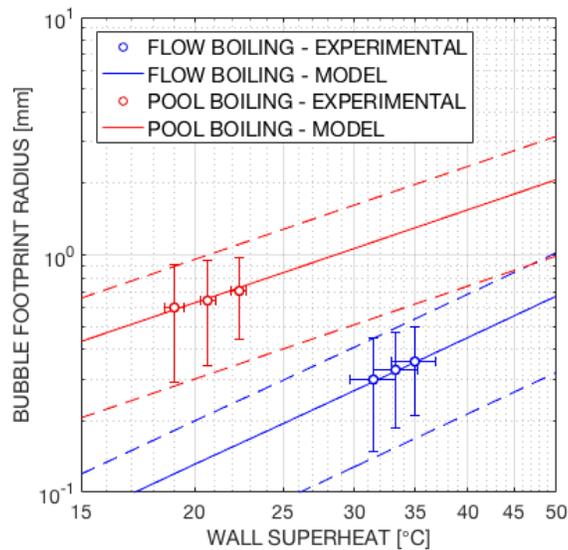

Figure S2.4. Measured footprint radius of individual bubbles as a function of wall superheat in flow boiling (blue dots) and pool boiling (red dots) experiments. Solid lines show the temperature scaling law, Eq. (6), based on the force balance model described in Ref. [3]. The error bar of bubble footprint radius is the standard deviation of the radius of all individual bubbles on the surface. The dashed lines represent the standard deviation of the bubble radius probabilistic density function, Eq. (8), calculated according to Eq. (7).

Mechanistic models to evaluate the bubble departure diameter are typically based on the force balance approach. Figure S2.5,

adapted from Ref. [3], shows the forces acting on a bubble growing on a surface, i.e., buoyancy ($F_b$), shear lift ($F_{sl}$), contact pressure ($F_{cp}$), hydrodynamic pressure ($F_h$), quasi-steady drag ($F_{qs}$), surface tension ($F_s$) and unsteady drag due to bubble growth ($F_{du}$). The bubble detaches from the boiling surface when there is a positive resultant force, $F_x$ or $F_y$, on the bubble.

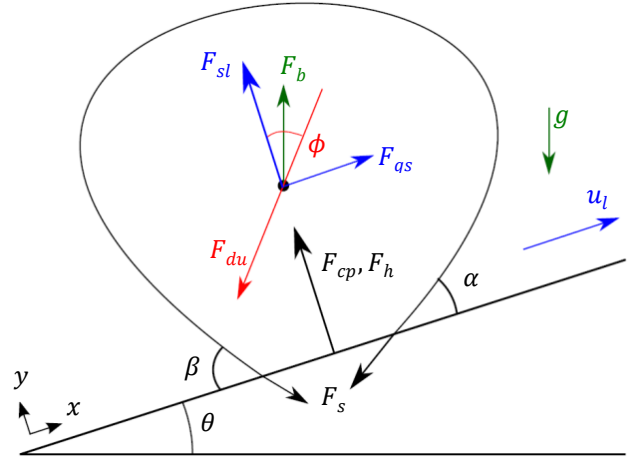

Figure S2.5. Individual forces on a bubble growing on the boiling surface (figure adapted from Mazzocco et al. [3]).

In flow boiling conditions, for $\theta = 90°$, detachment can happen by both departure ($\sum F_y > 0$ while $\sum F_x = 0$) or sliding ($\sum F_x > 0$ while $\sum F_y = 0$). However, the first one is the main mechanism in our operating conditions. Notably, the shear lift, $F_{sl}$, and the unsteady drag, $F_{du}$ are, by far, the dominant detaching and adhesive forces, respectively:

$$F_{sl} = \frac{1}{2} C_L \pi \rho_l w^2 R_b^2 \qquad (1)$$

$$F_{du} = \frac{1}{4} \pi \rho_l K^2 \qquad (2)$$

where $\rho_l$ is the liquid density, $C_L \sim 2.61$ is the lift coefficient, $R_b$ is the bubble radius and $w$ is the velocity of the fluid at the bubble center mass, which introduces a higher order dependence on the bubble radius. $K$ is given by:

$$K = \text{Ja}^* \sqrt{\eta_l} \left( C_{ml} \frac{1}{\sqrt{\text{Pr}_l}} + \chi C_{pb} \right) \qquad (3)$$

where $\eta_l$ and $\text{Pr}_l$ are the thermal diffusivity and the Prandtl number of the liquid, respectively, $C_{ml}$ and $C_{pb}$ are numerical constants, and $\chi$ is a parameter to account for the subcooling of water. The modified Jacob number, $\text{Ja}^*$, is given by

$$\text{Ja}^* = \frac{\rho_l C_{pl} \Delta T_{sat}}{\rho_v h_{fg}} \qquad (4)$$

where $C_{pl}$ is the specific heat of the liquid, $\rho_v$ is the vapor density, $h_{fg}$ is the latent heat, and $\Delta T_{sat}$ is the wall superheat. Briefly, when equating the shear lift and unsteady drag forces, one finds that, everything else being the same, the departure radius is proportional to the wall superheat

$$R_b \propto \Delta T_{sat}^\alpha \qquad (5)$$

A similar reasoning applies to pool boiling conditions, for $\theta = 0°$, where detachment happens when $\sum F_y$ becomes larger than 0.

Assuming that the bubble departure radius and the bubble footprint radius scale with the wall superheat in the same fashion, we can estimate the average bubble footprint radius $\langle R \rangle$ for high heat fluxes as:

$$\langle R \rangle = \langle R_0 \rangle \left( \frac{\Delta T_{sat}}{\Delta T_{sat,0}} \right)^{\alpha} \tag{6}$$

where the subscript 0 refers to experimental boiling conditions with the lowest heat flux and wall superheat, for which only isolated bubbles exist. They coincide with the points at ~19°C and ~31°C of wall superheat for the pool boiling and flow boiling data in Figure S2.4, respectively. Note that the available experimental data for isolated bubbles are in good agreement with the temperature scaling law, Eq. (6), shown as solid lines in Figure S2.4. In the same figure, the dashed lines represent:

$$\langle R \rangle \pm \sigma_{\langle R \rangle} \tag{7}$$

where $\sigma_{\langle R \rangle}$ is the standard deviation of the probabilistic density function

$$P(R) = 2\pi R c e^{-c\pi R^2} \tag{8}$$

where $c = 1/(4 \langle R \rangle^2)$, which is equal to

$$\sigma_{\langle R \rangle} = \sqrt{\langle R^2 \rangle - \langle R \rangle^2} = \sqrt{\frac{4-\pi}{\pi}} \langle R \rangle \tag{9}$$

Note that the experimental standard deviations are in good agreement with Eq. (7) (dashed lines in Figure S2.4), which supports the use of Eq. (8) as an input in the MC model. The cumulative density function (CDF) of this PDF, Eq. (8), is given by

$$C(R) = 1 - e^{-\frac{\pi R^2}{4 \langle R \rangle^2}} \tag{10}$$

When generating a bubble in the MC model, its radius is determined according to this CFD. Precisely, Eq. (10) is inverted to obtain

$$R = \sqrt{-\frac{4 \langle R \rangle^2}{\pi} \ln(1 - C(R))} \tag{11}$$

and $C(R)$ is replaced by a random number $q \in [0,1]$.